\documentclass[%
 reprint,
superscriptaddress,
 amsmath,amssymb,
 aps,
 floatfix,
 longbibliography,
]{revtex4-2}

\usepackage[utf8]{inputenc}
\usepackage[T1]{fontenc}

\usepackage[english]{babel}

\makeatletter
\adddialect\l@en\l@english

\makeatother

\usepackage{graphicx}
\usepackage{dcolumn}
\usepackage{bm}
\usepackage{booktabs}
\usepackage{tabularx}
\usepackage[version=4]{mhchem}
\usepackage{miller}
\usepackage{siunitx}
\usepackage{xcolor}
\usepackage[colorlinks=true,linkcolor=blue,citecolor=blue,urlcolor=blue]{hyperref}

\newcommand{\Et}{E_{\mathrm{t}}}
\newcommand{\Ec}{E_{\mathrm{C}}}
\newcommand{\Ef}{E_{\mathrm{F}}}
\newcommand{\Dit}{D_{\mathrm{it}}}
\newcommand{\Vtip}{V_{\mathrm{tip}}}

\newcommand{\fzero}{f_{0}}

\begin{document}

\title{Individual characterization of fast-responding trap states\\at the NO-annealed SiO$_2$/4H--SiC interfaces}

\author{Takahiro Ono}
\affiliation{Department of Advanced Materials Science, Graduate School of Frontier Sciences, The University of Tokyo, Kashiwa, Chiba, Japan}

\author{Mizuki Ohashi}
\affiliation{Department of Applied Physics, Graduate School of Engineering, The University of Tokyo, Bunky\={o}, Tokyo, Japan}

\author{Tomohiro Shigeno}
\affiliation{Department of Advanced Materials Science, Graduate School of Frontier Sciences, The University of Tokyo, Kashiwa, Chiba, Japan}

\author{Yutaro Uchida}
\affiliation{Department of Advanced Materials Science, Graduate School of Frontier Sciences, The University of Tokyo, Kashiwa, Chiba, Japan}

\author{Yuuki Yasui}
\affiliation{Department of Advanced Materials Science, Graduate School of Frontier Sciences, The University of Tokyo, Kashiwa, Chiba, Japan}

\author{Koji Kita}
\affiliation{Department of Advanced Materials Science, Graduate School of Frontier Sciences, The University of Tokyo, Kashiwa, Chiba, Japan}

\author{Yoshiaki Sugimoto}
\email{ysugimoto@k.u-tokyo.ac.jp}
\affiliation{Department of Advanced Materials Science, Graduate School of Frontier Sciences, The University of Tokyo, Kashiwa, Chiba, Japan}

\date{\today}

\begin{abstract}
Fast-responding trap states introduced by NO-annealing are suspected to limit the channel mobility of 4H--SiC MOSFETs, yet their microscopic characterization remains challenging because conventional electrical methods are spatially averaged and do not readily isolate such fast processes. Here, we visualize and analyze individual fast-responding trap states at the NO-annealed SiO$_2$/4H--SiC interface using the energy dissipation signal in frequency-modulation atomic force microscopy (FM--AFM), which selectively probes charge-exchange dynamics on sub-$\mu$s time scales. Ring-shaped dissipation patterns were observed in the NO-annealed sample but not in the control sample without NO-annealing, indicating that the detected states are associated with nitridation. Spectroscopic measurements were also performed to determine the dependence of energy dissipation on the tip bias and the tip-sample distance. Combined with finite-element electrostatic calculations, this analysis allowed us to determine trap energies relative to the Fermi level, $\Et-\Ef$, and revealed that the trap-energy distribution extends toward the interfacial conduction-band edge. These results provide microscopic evidence that NO-annealing generates fast-responding trap states near the SiO$_2$/4H--SiC interface.
\end{abstract}


\maketitle

\section{Introduction}

Wide-bandgap (WBG) semiconductors have attracted increasing attention as next-generation power-device materials, driven by the demand for higher efficiency and reliability in power conversion. Among WBG semiconductors, 4H--SiC is particularly promising because it combines a high critical breakdown electric field~\cite{godignon_breakdown_2024} with high thermal conductivity~\cite{qian_anisotropic_2017} and relatively weak anisotropy of electron mobility~\cite{kimoto_fundamentals_2014}. Moreover, a key processing advantage of SiC is that a high-quality SiO$_2$ gate dielectric can be readily formed by thermal oxidation, allowing metal--oxide--semiconductor (MOS) structures compatible with mature Si metal--oxide--semiconductor field-effect transistor (MOSFET) processing~\cite{li_investigation_2023,green_ultrathin_2001}.

Despite these advantages, the channel mobility in 4H--SiC MOSFETs is extremely low ($\approx 5~\mathrm{cm^2/Vs}$~\cite{yano_high_2000}) compared with the bulk electron mobility (1020--1200~$\mathrm{cm^2/Vs}$~\cite{kimoto_fundamentals_2014}). This degradation is generally attributed to a high density of electrically active defects remaining near the SiO$_2$/4H--SiC interface, including interface states and oxide-side near-interface traps. NO-annealing is commonly employed to improve the interface~\cite{vidarsson_improvement_2023,kil_anomalous_2020,rozen_increase_2008,chung_improved_2001}; however, even NO-annealed devices typically exhibit channel mobilities of only $\approx 30$--$35~\mathrm{cm^2/Vs}$ (for the typical cases of $\sim10^{16}~\mathrm{cm^{-3}}$ channel doping), which remain far below the values expected from bulk 4H--SiC~\cite{chung_improved_2001}.

Using high-frequency conductance analysis combined with theoretical modeling, Yoshioka \textit{et al.} pointed out the presence of fast-responding electrically active states associated with NO-annealing and further argued that interface-related traps with nanosecond-scale relaxation times, which are difficult to detect by conventional $C$--$V$ measurements (typically up to $\approx 1~\mathrm{MHz}$), can degrade channel mobility~\cite{yoshioka_generation_2012,yoshioka_characterization_2014}. However, the microscopic origin of these fast-responding states and their quantitative impact on device performance remain under debate, in part because electrical measurements are spatially averaged and the separation of interface states and near-interface oxide traps can be model dependent~\cite{wen_impact_2025,huang_effects_2025}.

In this study, we spatially resolve and evaluate individual fast-responding trap states at the NO-annealed SiO$_2$/4H--SiC interface using the energy dissipation signal in frequency-modulation atomic force microscopy (FM--AFM). We can clearly discuss the in-plane distribution of individual traps assumed to exist at the interface. In this dissipation-based scheme, trap charging and discharging within an oscillation cycle of cantilever produce a hysteresis in the electrostatic force and an associated increase in the dissipated energy, providing selectivity to trap states with sub-$\mu$s response times under our experimental conditions. Related dissipation-based approaches have previously been applied to isolated Au nanoparticles~\cite{tekiel_room-temperature_2013} and quantum dots~\cite{miyahara_quantum_2017}. Moreover, Cowie \textit{et al.} recently reported the visualization of interface states at the SiO$_2$/Si interface using a dissipation-based method~\cite{cowie_spatially_2024}.

\begin{figure}[t]
	\centering
	\includegraphics[width = 1\hsize]{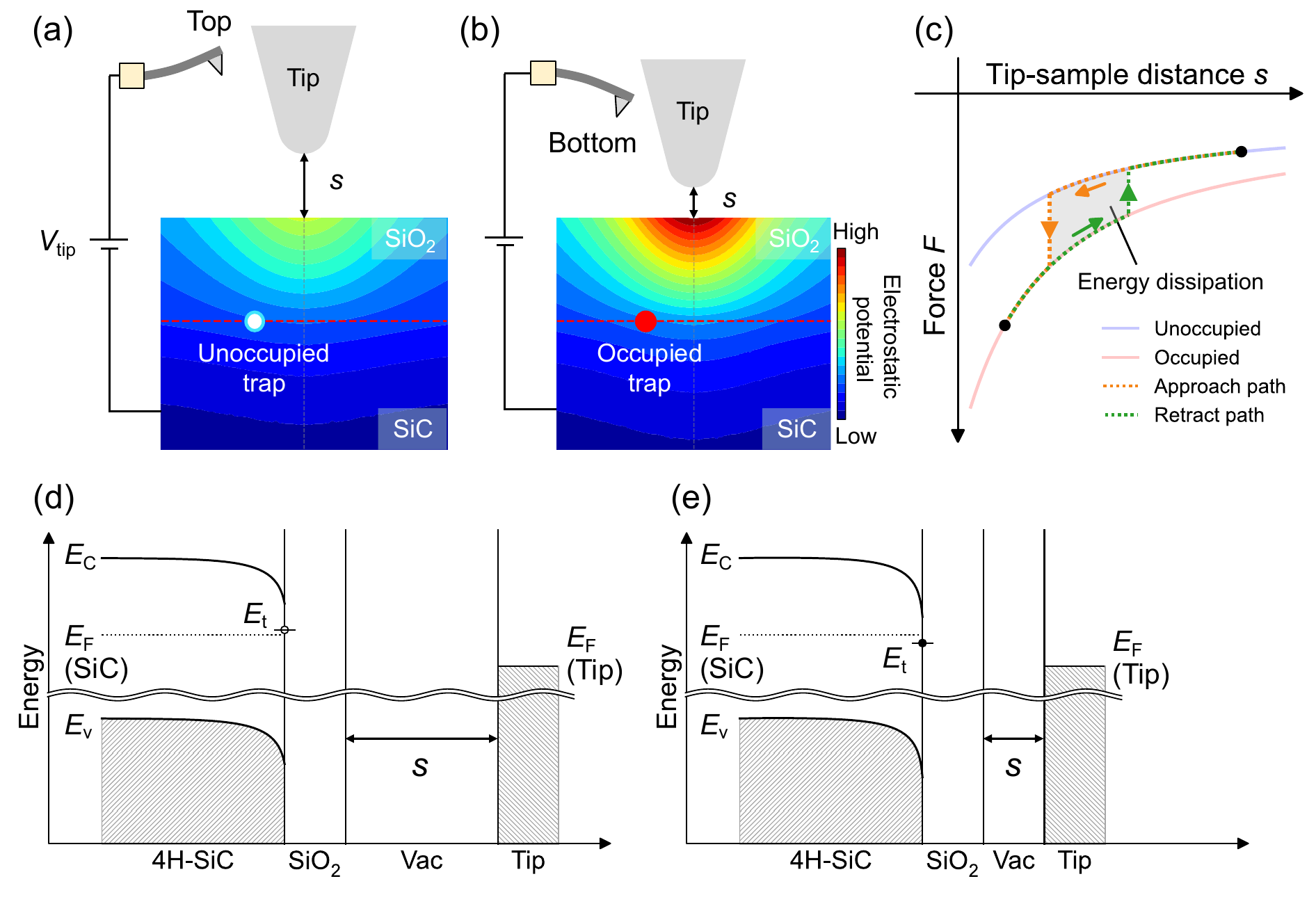}
	\caption{Schematic illustration of individual interface-trap detection by FM--AFM.
(a) Electrostatic potential at the top of the tip oscillation cycle.
(b) Electrostatic potential at the bottom of the cycle, where the higher potential induces trap occupation.
(c) Schematic force--distance curve showing dissipation caused by hysteretic trap charging and discharging.
(d) Band diagram of the SiC/SiO$_2$/vacuum/metallic-tip structure at positive tip bias with the tip farther from the surface.
(e) Corresponding band diagram with the tip closer to the surface.
Open and filled circles indicate unoccupied and occupied traps, respectively.}
	\label{fig1}
\end{figure}

Figure~\ref{fig1} summarizes the trap-detection scheme based on the FM--AFM energy dissipation signal. Here, $\Vtip$ denotes the tip bias with respect to the sample and $s$ the tip--sample distance. When a biased tip approaches the sample surface, the local electrostatic potential near the interface is modified, which can switch the charge occupancy of a trap [Figs.~\ref{fig1}(a) and (b)]. Figure~\ref{fig1}(c) schematically illustrates the force--distance curve when the trap occupancy changes during an oscillation cycle of the tip. The electrostatic force acting on the tip of cantilever depends on the trap charge state. Because charge transfer at the trap involves a finite relaxation time, occupancy switching leads to different forces on the approach and retract paths, resulting in a hysteresis loop. This hysteresis loop increases the energy dissipation and is detected as the excitation-voltage signal in FM-AFM. At a given $\Vtip$ and $s$, the switching condition is satisfied at a constant lateral distance between tip and trap, thereby visualizing individual traps as ring-shaped dissipation patterns. Figures~\ref{fig1}(d) and (e) show simulated band diagrams under positive tip bias. In this case, the SiC/SiO$_2$/vacuum/metallic-tip structure can be regarded as a MOS capacitor, in which the band bending near the interface is controlled by $\Vtip$ and $s$. As $s$ decreases, the trap level $\Et$ crosses the Fermi level $\Ef$, resulting in a change in trap occupancy.

\section{Results and Discussion}
\begin{figure}[t]
	\centering
	\includegraphics[width = 1\hsize]{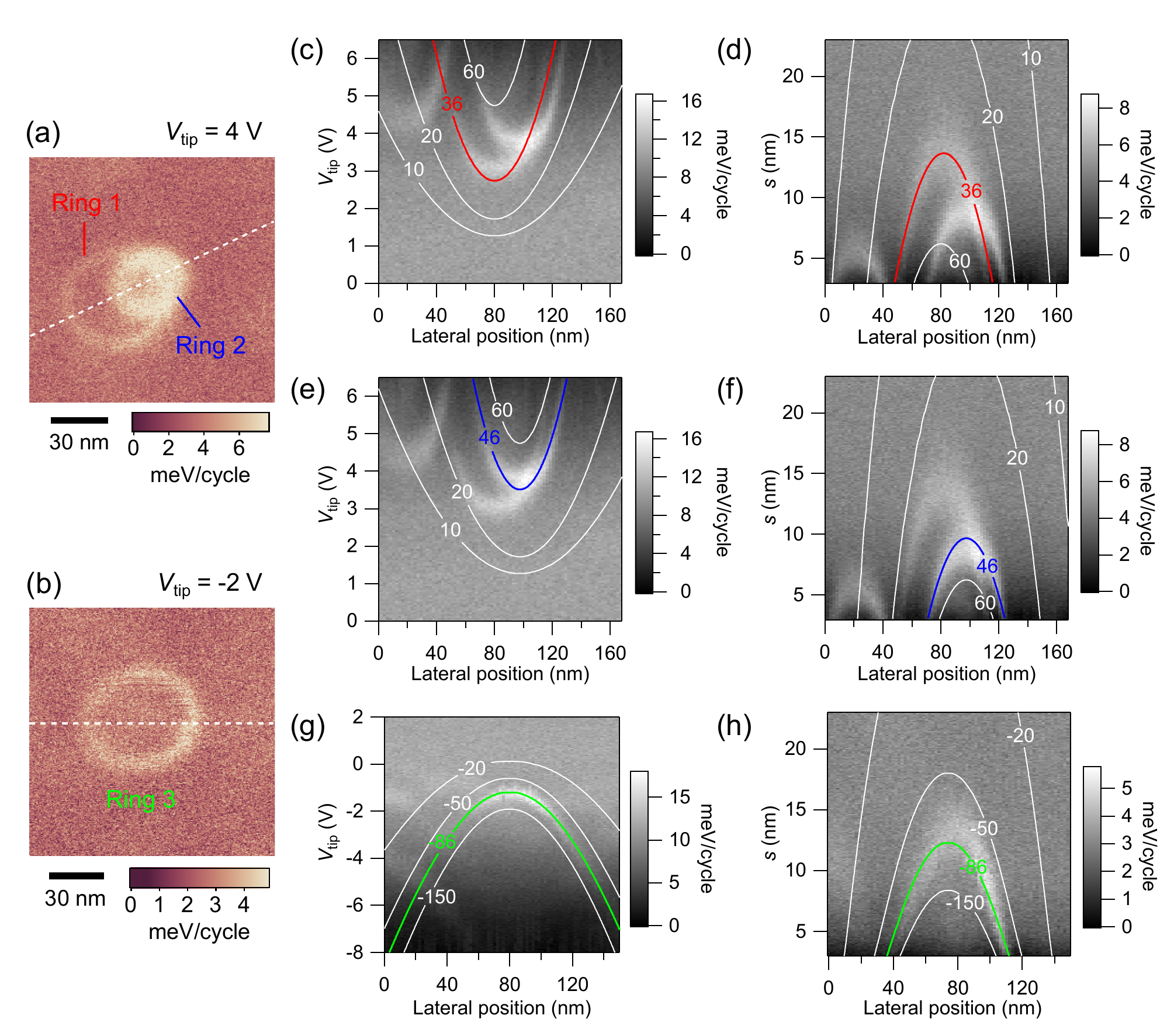}
	\caption{Energy dissipation imaging and dissipation spectroscopy of individual trap states at the NO-annealed SiO$_2$/4H--SiC interface.
(a) Dissipation image at $\Vtip=4~\mathrm{V}$, showing rings~1 and~2.
(b) Dissipation image at $\Vtip=-2~\mathrm{V}$, showing ring~3.
(c--h) Two-dimensional dissipation maps measured along the dashed lines in (a) and (b) while sweeping $\Vtip$ or $s$:
(c,d) ring~1, (e,f) ring~2, and (g,h) ring~3.
Overlaid curves indicate equipotential contours calculated by the finite element method, and the labels give the energy relative to $\Ef$ in meV.}
	\label{fig2}
\end{figure}

We performed FM--AFM energy dissipation measurements on a SiO$_2$/4H--SiC(0001) sample in which nitrogen was introduced near the interface by direct NO-annealing of the 4H--SiC(0001) substrate~\cite{uchida_study_2026}. Details of the sample preparation and FM--AFM measurements are given in the Experimental Section. Figures~\ref{fig2}(a) and (b) show dissipation images acquired at $\Vtip=4~\mathrm{V}$ and $-2~\mathrm{V}$, respectively. The ring-shaped contrast is attributed to charge exchange at individual traps, and each ring center indicates the lateral trap position. In contrast, no rings were observed for a control SiO$_2$/4H--SiC(0001) sample in which an approximately 5-nm-thick oxide was formed by dry oxidation without NO-annealing (Supporting Information), indicating that the observed trap states are associated with the nitridation process. Since it is expected that NO treatment would reduce traps, this is an interesting observation.

Rings~1 and~2 in Fig.~\ref{fig2}(a) have different radii, reflecting differences in trap energy. A smaller ring radius $R$ implies that a larger effective electrostatic potential is required to bring the trap level across the Fermi level, whereas a larger $R$ indicates that a smaller effective potential is sufficient. Figures~\ref{fig2}(c)--(h) show two-dimensional dissipation maps measured while sweeping $\Vtip$ and $s$ along the dashed lines in Figs.~\ref{fig2}(a) and (b), which pass through the centers of the rings corresponding to individual trap positions. The parabolic bright ridges in these maps correspond to individual trap states, and the $\Vtip$ and $s$ dependences of the ridge opening reflect the trap energy.

To quantify the trap energies, we calculated the electrostatic potential distribution using the finite element method and simultaneously fitted the calculated $R(\Vtip)$ and $R(s)$ curves to the experimental data (see Supporting Information for details). The electrostatic potential that simultaneously reproduces the observed $\Vtip$ and $s$ dependences was identified with the energy offset between the trap level and the Fermi level in 4H--SiC, $\Et-\Ef$. From this analysis, we obtained $\Et-\Ef = 36$, 46, and $-86~\mathrm{meV}$ for rings~1--3, respectively.

\begin{figure}[t]
	\centering
	\includegraphics[width = 1\hsize]{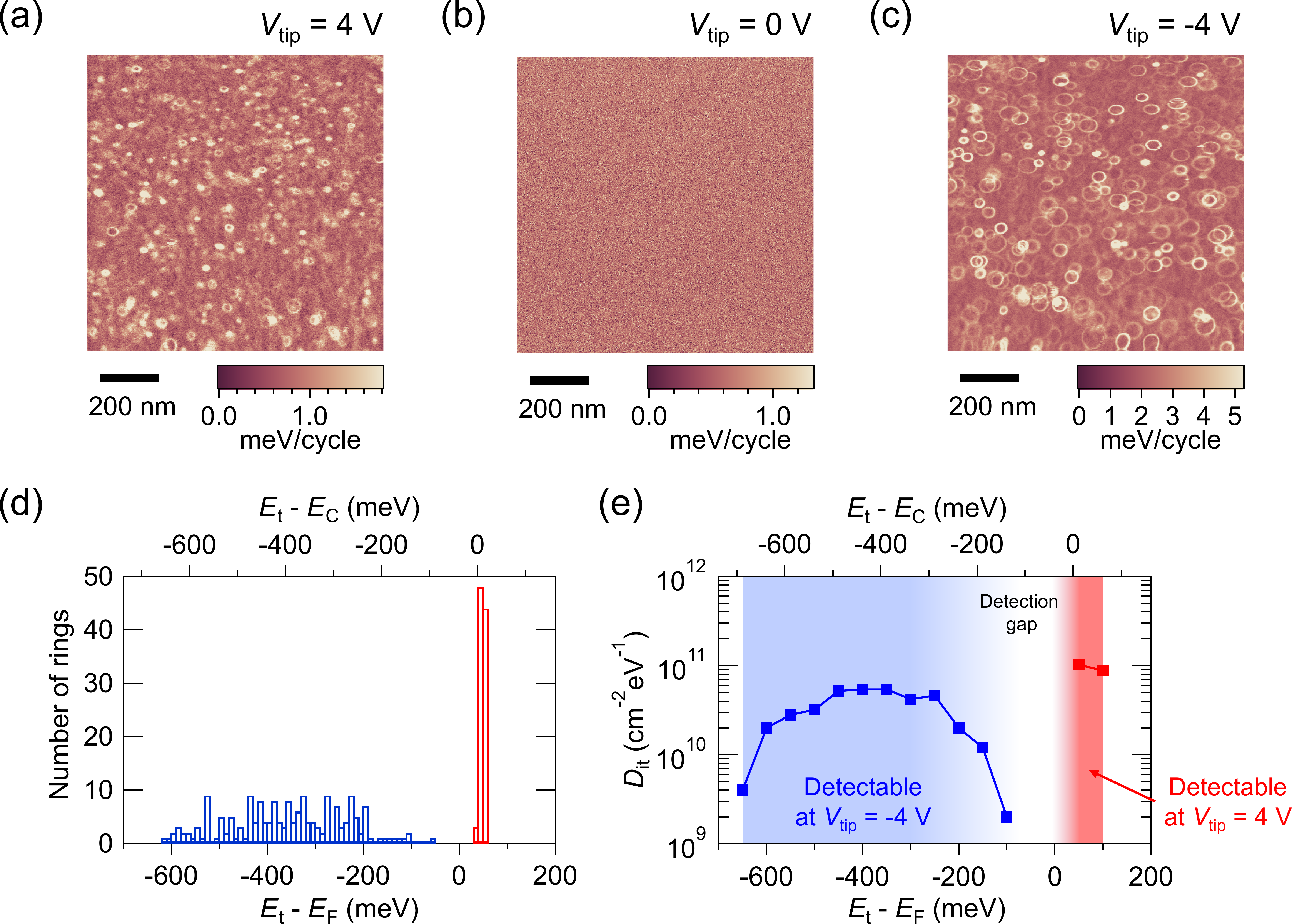}
\caption{(a--c) Dissipation images acquired over a $1\times1~\mu\mathrm{m}^2$ area at $\Vtip=4~\mathrm{V}$, $0~\mathrm{V}$, and $-4~\mathrm{V}$, respectively.
(d) Ring-count distribution as a function of $\Et-\Ef$ (bottom axis) and the corresponding $\Et-\Ec$ values on the top axis, obtained from the detected ring radii using a bin width of 10~meV. Red and blue bins correspond to (a) and (c), respectively.
(e) Energy distribution of $\Dit$ derived from (d). Red and blue shaded regions indicate the experimentally accessible energy windows at $\Vtip=4~\mathrm{V}$ and $-4~\mathrm{V}$, respectively.}
	\label{fig3}
\end{figure}

To investigate the energy distribution of detectable trap states, we analyzed dissipation images acquired at different tip biases [Figs.~\ref{fig3}(a)--(c)]. A custom deep-learning-based program automatically detected ring patterns and extracted the ring radius $R$~\cite{czarnecki_hydrogen_2025, paszke_pytorch_2019}, yielding 95 rings at $\Vtip=4~\mathrm{V}$ [Fig.~\ref{fig3}(a)] and 183 rings at $\Vtip=-4~\mathrm{V}$ [Fig.~\ref{fig3}(c)]. The extracted radii were then converted into trap-energy distributions using electrostatic potential simulations [Fig.~\ref{fig3}(d)]. Figure~\ref{fig3}(e) shows the resulting energy distribution of the interface trap density $\Dit$ for $\Vtip=4~\mathrm{V}$ and $-4~\mathrm{V}$. The extracted $\Dit$ is on the order of $10^{10}$--$10^{11}~\mathrm{cm^{-2}eV^{-1}}$, which is comparable to that of fast interface states reported for similar NO-annealed samples~\cite{yoshioka_generation_2012,yoshioka_characterization_2014}. 
Assuming that the interfacial band bending is laterally uniform in the absence of an external electric field, the energy distribution can also be expressed relative to the conduction-band edge at the interface, $\Ec$. Here, $\Ec$ denotes the interfacial conduction-band edge after band bending rather than the bulk band edge.
 In this representation, $\Dit$ values are larger near $\Ec$ than in  deeper energy regions. That is consistent with previous electrical studies~\cite{zhai_electrical_2020, fiorenza_impact_2024, huang_effects_2025} and indicates that the observed states are dominated by shallow acceptor-like electron trap states near the interfacial conduction-band edge.
 
We note that the reduced counts in the range $-250 < \Et-\Ef < 50~\mathrm{meV}$ do not imply an intrinsic absence of trap states, but are most likely a methodological limitation. Because the dissipation contrast decays with increasing radius, trap states that respond only at very large ring radii under $\Vtip=\pm4~\mathrm{V}$ can produce signals too weak to exceed the detection threshold. 

The ring patterns were observed only for the NO-annealed sample and were absent in the O$_2$-annealed sample, indicating that the detected trap states are induced by nitridation. Because measurable dissipation requires charge switching of the trap within one oscillation cycle of the tip, the experimentally accessible relaxation time is estimated to be shorter than $1/(2\fzero) = 1.8~\mu\mathrm{s}$, where the cantilever's resonance frequency $\fzero = 284~\mathrm{kHz}$. We could also observe the trap states in the cantilever’s second resonance mode with $f = 1.76~\mathrm{MHz}$ (Supporting Information).
These results indicate that the nitrogen-induced interface states respond well below 1~$\mu$s, consistent with previous reports~\cite{yoshioka_generation_2012,yoshioka_characterization_2014}. 
Our approach, which can analyze individual traps, does not rely heavily on modeling in contrast to the $\Dit$ estimation from device characteristics based on a modeling with presumed equivalent circuits. Therefore, we can confidently conclude that NO-annealing generates fast-responding trap states near the SiO$_2$/4H--SiC interface. A plausible microscopic explanation is that nitrogen incorporated at the interface forms electrically active defects; one candidate is an NSi$_3$-like configuration~\cite{xu_atomic_2014}, although more complex nitrogen-related interfacial defects may also contribute. These results suggest that, although NO-annealing is widely used to improve the SiO$_2$/4H--SiC interface, it can simultaneously generate fast-responding trap states near the conduction-band edge. These traps may affect the device's high-frequency operation.

\section{Conclusion}
In summary, we directly observed fast-responding trap states near the SiO$_2$/4H--SiC interface using FM--AFM. Individual trap states were spatially separated and visualized as ring patterns in dissipation images. By comparing the dissipation maps with electrostatic potential simulations, we identified the energy levels of individual trap states in the nanoscale region. The presence of a high density of fast-responding trap states extending toward $\Ec$ suggests that NO-related defects provide electrically active states very close to the conduction-band edge. Our approach should be broadly applicable to the analysis of trap states in various power semiconductor and quantum-sensing devices.

\section{Experimental Section}

\subsection{Sample preparation}

The sample structure was a 5-nm-thick SiO$_2$ layer on a 5-$\mu$m-thick nitrogen-doped ($n$-type) 4H--SiC(0001) epitaxial layer with $N_\mathrm{D} \approx 1 \times 10^{16}~\mathrm{cm^{-3}}$. After HF cleaning, the sample was directly annealed at 1150$^\circ$C for 120~min in an NO/N$_2$ ambient ($\mathrm{NO}/\mathrm{N}_2=1/2$) to form a SiO$_2$ layer with nitrogen incorporated mostly near the interface~\cite{uchida_study_2026}. The nitrogen content in the bulk part of the film was confirmed to be only 1 at.\%. After loading the sample into the vacuum chamber, FM--AFM measurements were performed without any degassing.

\subsection{FM--AFM measurements}

All AFM measurements were performed at room temperature in an ultrahigh-vacuum chamber ($\sim 10^{-11}~\mathrm{Torr}$) using a custom-built AFM based on optical interferometry and operated in the frequency-modulation (FM) mode. In this mode, the energy dissipation was evaluated from the excitation voltage required to maintain a constant cantilever oscillation amplitude.

Commercial Pt-coated Si cantilevers were used as force sensors. The cantilevers had a resonance frequency of $\fzero = 284$~kHz, a spring constant of $k = 19$~N/m, and a quality factor of approximately $2\times10^{4}$. Before the measurements, the cantilevers were cleaned by Ar$^+$ ion sputtering ($0.6~\mathrm{keV}$, $3\times10^{-7}~\mathrm{Torr}$, 25~min). This sputter-cleaning process removes the native oxide and other contaminants from the tip apex, thereby maintaining good tip conductivity during the measurements. The oscillation amplitude was set to $A = 500$~pm.

\begin{acknowledgments}
This work was supported by JST SPRING Grant No.~JPMJSP2108, JST FOREST Program Grant No.~JPMJFR203J, and JSPS KAKENHI Grant Nos.~25K22215, 26K01386, 26H01330, and 24H00308. Y.~S. acknowledges support from the Precise Measurement Technology Promotion Foundation (PMTP-F) and the Iwatani Naoji Foundation.
\end{acknowledgments}


\bibliography{reference}

\end{document}